\documentstyle[preprint,tighten,aps]{revtex}
\begin{document}

\title{\bf Vortex Pair Production and Decay of a 2-D Supercurrent
\\ by a Quantum Field Theory Approach }

\author{ Roberto Iengo \\ {\it International School for Advanced
Studies \\ Via Beirut 4, 34014 Trieste (Italy) \\ INFN -- Sezione di
Trieste } }

\author{ Giancarlo Jug \\ {\it INFM -- Istituto di Scienze
Matematiche, Fisiche e Chimiche \\ Universit\`a di Milano a Como, Via
Lucini 3, 22100 Como (Italy) \\ INFN -- Sezione di Pavia } }

\maketitle

\vskip 1.0truecm

\abstract{ We investigate the phenomenon of the decay of a
supercurrent through homogeneous nucleation of vortex-antivortex
pairs in a 2-D like superconductor or superfluid by means of a
quantum electrodynamic formulation for the decay of the 2-D vacuum.
The case in which both externally-driven current and Magnus force are
present is treated exactly, taking the vortex activation energy and
its inertial mass as independent parameters.  Quantum dissipation is
included through the formulation introduced by Caldeira and Leggett.
The most relevant consequence of quantum dissipation is the
elimination of the threshold for vortex production due to the Magnus
force.  In the dissipation-dominated case, corresponding formally to
the limit of zero inertial mass, an exact formula for the pair
production rate is given.  If however the inertial mass is strictly
zero we find that vortex production is inhibited by a quantum effect
related to the Magnus force.  The possibility of including vortex
pinning is investigated by means of an effective harmonic potential.
While an additional term in the vortex activation energy can account
for the effect of a finite barrier in the direction perpendicular to
the current, pinning along the current depresses the role of the
Magnus force in the dissipation-dominated dynamics, except for the
above-mentioned quantum effect.  A possible description of vortex
nucleation due to the combined effects of temperature and
externally-driven currents is also presented along with an evaluation
of the resulting voltage drop.  }

\pacs{PACS numbers:  74.60.Ge, 74.20.-z, 03.70.+k }

\vfill \newpage

\section{ Introduction }
\renewcommand{\theequation}{1.\arabic{equation}}
\setcounter{equation}{0}

\noindent

The layered structure and high critical temperature of the oxide
superconductors has lead to much renewed interest in the behaviour of
these materials under applied magnetic fields.  Of particular
interest are the statics and dynamics of the magnetic vortices in the
presence of pinning impurity centers, where issues such as the
formation of a glassy vortex phase \cite{geilo} and those connected
to transport properties such as flux-creep \cite{blatter} and vortex
tunneling \cite{ivlev,thouless,stephen} have been vivaciously debated
in the recent literature.  Similar questions of vortex dynamics have
attracted the interest of researchers in the field of superfluidity
for the past three decades \cite{tilley,donnelly}.

An issue that appears to lend itself to full analytic treatment by
means of a path-integral formulation is the quantum tunneling of a
vortex trapped by a pinning center in a 2-D superfluid or
superconductor structure \cite{thouless,stephen}.  This problem has
been investigated both with and without the inclusion of quantum
dissipation and discussions that give full weight to the inertial
mass of the moving vortex can be found in the literature alongside
treatments \cite{stephen} in which the inertial mass is taken to be
negligible.  The physics of the problem is captured by semi-classical
evaluations of the tunneling rate and discussion has centered on the
effects of dissipation and pinning in opposing the stabilising
influence of the Magnus force on the classical orbits of a vortex.
Mathematically, and in the classical framework for a single vortex
moving in a supercurrent at relatively low velocity, the equation of
motion in the 2-D plane reads:

\begin{equation} m\ddot{\bf q}=-{\nabla}V({\bf q})-e\dot{\bf
q}{\times}{\bf B}-\eta\dot{\bf q} \label{classic} \end{equation}

\noindent Here $m$ is the inertial mass of the vortex of topological
charge $e=\pm2\pi$, treated as an single, point-like particle of 2-D
coordinate ${\bf q}(t)$.  Also, $V({\bf q})=V_p({\bf q})+V_v({\bf
q},{\bf J})$ is the phenomenological pinning potential plus the
electric-like potential due the supercurrent ${\bf J}$, namely
$-{\nabla}V_v=e{\bf E}={\times}e{\bf J}$ (the latter being a compact
notation for the 2-D dual of a vector:  $({\times}{\bf
J})_{\mu}={\epsilon}_{\mu\nu}J_{\nu}$).  Finally ${\bf B}={\hat{\bf
z}}d\rho_s^{(3)}$ is the magnetic-like field of the Magnus force
\cite{tilley,aoth,thaoni,gait} acting on the vortex in the plane of
the film of thickness $d$ and directed along the vector $\hat{\bf z}$
orthogonal to it and associated with a superfluid component having
3-D number density $\rho_s^{(3)}$; and $\eta$ is the phenomenological
friction coefficient.  We neglect in this work the additional
magnetic field intrinsic to the vortex since, as discussed by Clem
\cite{clem}, the flux quantum is homogeneously spread over a very
large distance for a single flux vortex in a 2-D superconductor.  The
quantum-mechanical version of Eq.  (\ref{classic}) is attained via
the Feynman path-integral transposition in which the dissipation is
treated quantistically through the formulation due (for the ohmic
case) to Caldeira and Leggett \cite{calleg}.

A situation closely related to the quantum tunneling of a vortex in
the presence of dissipation is the {\it decay of a supercurrent},
whether in a charged superconductor or in a neutral superfluid, due
to the spontaneous homogeneous nucleation of vortex-antivortex pairs
from the ``vacuum'' of the perfect superfluid wavefunction, in the
presence of an external uniform field.  In this case the field
corresponds -- in its electric part -- to the dual of the external
supercurrent ${\bf J}$, which may be kept fixed in the system if a
voltage drop is measured at the edge of the superconducting sample,
and -- in its magnetic part -- to the Magnus force field term
\cite{tilley,donnelly} ${\bf B}$.  This situation is the subject of
the present article, in which we wish to show that the lifetime
${\Gamma}^{-1}$ of the supercurrent can be calculated explicitely in
terms of the phenomenological parameters of the material by means of
a straightforward application of the field-theoretic relativistic
formalism leading to the decay rate of the vacuum in scalar quantum
electrodynamics.  Indeed, the excitations of a relativistic quantum
field encode in a rather natural way the particle-antiparticle
quantum fluctuations of the vacuum and therefore the
particle-antiparticle creation.  This formalism, independently
already proposed by Ping Ao \cite{ao}, albeit for the simplest of the
cases treated herewith, allows us to evaluate $\Gamma$ for a number
of vortex dynamics cases relevant to experimentally accessible
situations.  With reference to the work of Ping Ao \cite{ao}, it
would appear that a main difference with the present work is the fact
that the effects of the Magnus force, of dissipation and/or of
pinning were not included in the resulting formula for the vortex
pair production rate.  For all the above-mentioned cases, the central
technical ingredient of our formulation for the zero-temperature
dynamics is an effective screened interaction and the resulting
quadratic form of the effective-particle Lagrangian for the Feynman
path-integral formulation of vacuum decay.

Indeed, our article is organised as follows.  In Section II we sketch
some basic facts about the vortex dynamics and introduce our quantum
field theory (QFT) formulation.  This is expanded in Section III,
where we treat in detail the situation in which dissipation and
pinning are both absent, with the supercurrent inducing a
vortex-antivortex pair production in the presence of the homogeneous
fields of the supercurrent and of the Magnus force.  Two
phenomenological parameters are introduced to describe the isolated
vortex:  its nucleation energy ${\cal E}_0$, suitably renormalised
(e.g.  to include screening effects by the external current or by the
vortex plasma above the Kosterlitz-Thouless (KT) transition and
pinning by impurities), and the vortex inertial mass $m$.  We
consider $m$ and ${\cal E}_0$ to be independent phenomenological
parameters so as to account for the issue, debated in the recent
vortex-dynamics literature \cite{stephen,niaoth}, where the inertial
mass $m$ may be taken as negligible (but see also \cite{blatter}).
It is seen that the Magnus force raises a threshold for the magnitude
of the supercurrent (or, the ``electric'' field) in which the pair
production may occur.  This was also qualitatively observed in
\cite{thouless,stephen}).  In Section IV we include the effects of
quantum dissipation through the formulation of Caldeira and Leggett
\cite{calleg}, in which the drag force experienced by the moving
vortex is mimicked by a linear coupling to the coordinates of an
infinite set of harmonic oscillators.  The main qualitative effect is
seen to be the suppression of the threshold for pair production, as
the Landau-level stability of the vortex in the field of the Magnus
force is destroyed by any infinitesimal amount of dissipation.  The
formula for the production rate $\Gamma$ is worked out in detail for
the dissipation-dominated case, that is in the limit
${\gamma}=m/{\cal E}_0{\rightarrow}0$.  We deal mainly with the
ohmic-case, to then show that sub- and super-ohmic cases can be
treated in a similar way to yield a less pronounced pair-production
phenomenon.  In Section V we discuss the possibility of including the
effects of pinning centres.  A finite pinning barrier in the
direction perpendicular to the current essentially results in a
contribution to the activation energy ${\cal E}_0$, however pinning
in the direction of the current has the consequence of suppressing
the effect of the Magnus force in the dissipation-dominated
dynamics.  We also discuss the extreme case where pinning can be
mimicked via an (unbounded) anisotropic harmonic force.  For large
enough separation, this corresponds to a confining linear potential
in our relativistic formulation and this may mimick a situation in
which a uniform pinning pressure is exerted.  The result is that, as
soon as a component of the force in the direction perpendicular to
the current is present, a threshold for vortex production appears.
Section VI contains a more tentative discussion on how to implement
the results obtained by the QFT approach in order to evaluate the
vortex-induced voltage drop in the direction of the current across
the sample, including also the effect of vortices which may be
activated by thermal fluctuations.  Finally, in the Appendix we show
that some of the qualitative features observed within the QFT
approach can also be seen to agree with a study of the classical
equations of motion for a single vortex.

\vskip 1.0truecm

\section{ Vortex dynamics:  a quantum field theory approach }
\renewcommand{\theequation}{2.\arabic{equation}}
\setcounter{equation}{0}

We begin this Section by briefly recalling the description of vortex
dynamics based on the time-dependent Landau-Ginzburg (TDLG)
formulation for the superfluid system in the presence of vortex
solutions and of a supercurrent.  For more extensive discussions,
also of related theoretical issues, the reader is referred for
instance to the articles of the Seattle group \cite{aoth},
\cite{niaoth,thaoni} and that of Lee and Fisher \cite{leefish} (see
also \cite{gait}).  For a superconductor, the TDLG effective action
for the wavefunction $\psi(x,y,t)=\sqrt{\rho_s}e^{i{\theta}(x,y,t)}$
corresponds to the Lagrangian (in the units in which ${\hbar}=1$ and
the carriers' charge $e^*=1$):

\begin{equation} {\cal L}=-\rho_s\dot{\theta}-\frac{1}{2m_0}\rho_s
\left ( \nabla \theta-{\bf A} \right )^2 +
f(\rho_s,\dot{\rho_s},\nabla\rho_s) \label{lagr} \end{equation}

\noindent Here $m_0$ is the effective mass of the carriers, $\rho_s$
the 2-D superfluid density and $f$ contains the additional, phase
independent, terms of the LG expansion in $|\psi|$ and its
derivatives.  In the following, we shall make the approximation of
constant $\rho_s$, thus dropping
$f(\rho_s,\dot{\rho_s},\nabla\rho_s)$, and treat the vortices as
point-like objects.  The external supercurrent is related to the
external vector potential ${\bf A}$ via

\begin{equation} {\bf J}=\frac{{\delta}{\cal L}}{{\delta}{\bf
A}}=\frac{\rho_s}{m_0}(\nabla\theta-{\bf A}) \end{equation}

\noindent Our situation consists in assuming that ${\bf J}$ is fixed
by some external device ``driving'' a constant current which we take
to be uniform throughout the sample.  We are then interested in
estimating the longitudinal potential drop and thus the ohmic
resistance (in general current-dependent) due to the production and
motion of the vortices.  We implement the requirement of fixed
current ${\bf J}$ by adding the term ${\bf A}{\cdot}{\bf J}$ to the
Lagrangian and treating ${\bf A}$ as a Lagrange multiplier:

\begin{equation} {\cal L}=-\rho_s\dot{\theta}-\frac{1}{2m_0}\rho_s
\left ( \nabla\theta-{\bf A} \right )^2-{\bf A}{\cdot}{\bf J}
\end{equation}

\noindent Then ${\delta}{\cal L}/{\delta}{\bf A}=0$ fixes the
current:  ${\bf J}=\frac{\rho_s}{m_0}\left(\nabla\theta-{\bf
A}\right)$; substituting back and disregarding a constant term
$\frac{m_0}{2\rho_s}J^2$, we get

\begin{equation} {\cal L}=-\rho_s\dot{\theta}-\nabla\theta{\cdot}{\bf
J} \end{equation}

\noindent A vortex is introduced as a configuration of the phase
$\theta({\bf r},t)$ having non-trivial topology and such that
$\nabla\theta$ has formally a non-vanishing rotational (a scalar, in
2D):

\begin{equation} \nabla{\times}\nabla\theta=\rho_v({\bf r})
\label{rhov} \end{equation}

\noindent In the presence of such singular vortex configurations with
topological charges $e_n={2\pi}n$ (where, in the following, $n=\pm1$)
and treating the vortices as point-like particles, we can rewrite the
action in terms of a vortex density $\rho_v({\bf
r})=\sum_ne_n{\delta}({\bf r}-{\bf r}_n)$ and vortex current density
${\bf J}_v({\bf r})=\sum_ne_n\dot{\bf r}_n{\delta}({\bf r}-{\bf
r}_n)$, with ${\bf r}_n(t)$ the vortex position.  The result is as
follows

\begin{eqnarray} {\cal S}&=&\int dt d^2r {\cal L}=\int dt d^2r \left
(-\rho_v V_v - {\bf J}_v{\cdot}{\bf A}_v \right ) \nonumber
\\ &=&\int dt \left \{ -\sum_n e_n V_v({\bf r}_n) - \sum_n e_n
\dot{\bf r}_n{\cdot}A_v({\bf r}_n) \right \} \label{tdlg}
\end{eqnarray}

\noindent where we have introduced the vortex fields ${\bf A}_v$ and
$V_v$ through $\nabla{\times}{\bf A}_v=B=\rho_s$ and ${\bf
E}=-{\nabla}V_v={\times}{\bf J}$.  One can formally obtain the Magnus
force term $-{\bf J}_v{\cdot}{\bf A}_v$ from the
$-\rho_s\dot{\theta}$ term in the TDGL expansion (\ref{lagr}), in a
shorthand way and disregarding possible subtle questions related to
boundary conditions (for a more thorough approach see for instance
\cite{aoth,thaoni,gait,leefish}).  If $\theta({\bf r})=\sum_n e_n
\theta \left ( {\bf r}-{\bf r}_n(t) \right )$, then
$\dot{\theta}=-\sum_n e_n \dot{\bf r}_n{\cdot}{\nabla}\theta({\bf
r}-{\bf r}_n)$.  If we denote by ${\Delta}^{-1}$ the inverse of the
Laplacian operator, then
$\dot{\theta}={\epsilon}_{\mu\nu}{\Delta}^{-1}{\partial}_{\nu}J_{v\mu}$,
where the vortex current is $J_{v\mu}=\sum_n e_n
\dot{x}_{n\mu}{\delta}({\bf r}-{\bf r}_n)$ and we have written, from
Eq.  (\ref{rhov}), ${\partial}_{\mu}\theta({\bf r}-{\bf
r}_n)=-{\epsilon}_{\mu\nu}{\Delta}^{-1} {\partial}_{\nu}{\delta}({\bf
r}-{\bf r}_n)$.  Therefore

\begin{equation} \int d^2r \rho_s\dot{\theta}=\int d^2r {\bf
A}_v{\cdot}{\bf J}_v \end{equation}

\noindent having formally put
$A_{v\mu}=-{\epsilon}_{\mu\nu}{\Delta}^{-1}{\partial}_{\nu}\rho_s$.
Also, it is straightforward to show that

\begin{equation} \int d^2r {\bf J}{\cdot}{\nabla}\theta=-\int d^2r
{\nabla}V_v{\cdot} ({\times}{\nabla}\theta)=\int d^2r V_v\rho_v
\end{equation}

\noindent Adding a kinetic energy term $\sum_n \frac{1}{2}m\dot{\bf
r}_n^2$ to the Lagrangian, Eq.  (\ref{tdlg}) shows that on the vortex
acts a Lorentz-like force linked to a magnetic-like field ${\bf B}$
of intensity $\rho_s$.  We also recover the classical equation of
motion (\ref{classic}) for the vortex:  the electric part of the
field acting on the vortex is associated with the dual vector of the
externally-applied supercurrent ${\bf J}$.  The picture is therefore
quite reminiscent of the situation for scalar two-space-dimensional
quantum electrodynamics (QED).

In practice, the ``electric field'' ${\bf E}={\times}{\bf J}$ will be
really uniform only on some scale for which the current ${\bf J}$ is
uniform.  For scales less than that, the vortices themselves will
alter the uniformity of the current.  In the language of QED, a
vortex at position ${\bf r}={\bf r}_1$ near an antivortex at position
${\bf r}={\bf r}_2$ will also feel -- beside the background
electrostatic potential $V_v({\bf r}_1)$ corresponding to the uniform
electric field -- the two-dimensional electrostatic attraction due to
the antivortex

\begin{equation} {\Delta}V_v({\bf r}_1)=\frac{e^2\rho_s}{2{\pi}m} \ln
\frac{|{\bf r}_1-{\bf r}_2|}{a} \end{equation}

\noindent where $a$ is some scale which we can conveniently take to
represent the vortex size.  The above vortex-antivortex
``electrostatic'' interaction gives rise to an additional barrier,
which can be taken into account \cite{minn,ivlev} by an additive
renormalization term in the activation energy.  Following Minnhagen
\cite{minn} (see also \cite{ivlev}), we make use of the fact that the
barrier maximum occurs for $|{\bf r}_1-{\bf
r}_2|{\sim}\frac{\rho_s}{m_0E}=\frac{\rho_s}{m_0J}$ thus contributing
a term to be included in the renormalization of the activation energy
of a vortex ($e_1=2\pi$)

\begin{equation} {\cal E}_{0R}={\cal
E}_0-\frac{\pi\rho_s}{\tilde{\epsilon}m_0} \ln \left (
\frac{m_0J}{\rho_sa} \right ) + {\cdots} \label{ren1} \end{equation}

\noindent Notice the appearance of the dielectrict constant
$\tilde{\epsilon}=\tilde{\epsilon}(T)$ which keeps into account the
polarization due to the thermal effects below the KT transition.
Above the KT transition, the ``electrostatic'' interaction is
screened by the KT screening length ${\lambda}_{KT}$ and the
renormalized activation energy will be, for
$J{\ll}\rho_s/m_0{\lambda}_{KT}$

\begin{equation} {\cal E}_{0R}={\cal E}_0-\pi\frac{\rho_s}{m_0}\ln
\left ( a/{\lambda}_{KT} \right ) + {\cdots} \label{ren2}
\end{equation}

\noindent In the following we will lump these renormalization effects
in the definition of the activation energy, which we call simply
${\cal E}_0$.

The central issue of the present work is that the ``electric'' field
term $\rho_vV_v$ in Eq.  (\ref{tdlg}) generates an instability in the
system, which will have a tendency of producing vortex-antivortex
pairs.  We propose to describe pair creation through the formalism of
relativistic QED.  In this formalism, pairs are created out of the
vacuum in the presence of a constant electromagnetic field, much as
in our electromagnetic-analogue situation for the vortex dynamics.
Further terms will have to be added to the action (\ref{tdlg}), which
describes only the interaction with the electromagnetic field, in
order to account for the vortex motion.  There will be, first of all,
the energy ${\cal E}_0$ of the isolated vortex at rest in the absence
of the external current.  This we take as an effective renormalised
activation energy ${\cal E}_{0R}$ to include the electrostatic
screening energy ${\Delta}{\cal E}$ as well as the average energy
required to overcome possible pinning potential barriers.  We
consider, to keep the discussion as general as possible, also a
kinetic energy term $\frac{m}{2}\dot{\bf r}^2$, where $m$ is the
possible inertial mass of the vortex.  In the relativistic formalism,
which we employ, the total ``free particle'' energy for a vortex
having momentum ${\bf p}$ would then be

\begin{equation} {\cal E}=\sqrt{{\cal E}_0^2+\frac{1}{\gamma}{\bf
p}^2}{\simeq}{\cal E}_0+\frac{1}{2m}{\bf p}^2+ {\cdots}={\cal
E}_0+\frac{m}{2}\dot{\bf r}^2+{\cdots} \label{rel} \end{equation}

\noindent Notice that in (\ref{rel}) the parameter
$\frac{1}{\gamma}={\cal E}_0/m$ plays the role of the square of the
speed of light $c$, in that the limit ${\gamma}{\rightarrow}0$
reproduces the non-relativistic ($c{\rightarrow}{\infty}$) expression
for the kinetic energy.  We imagine that the vortices will have a
rather small kinetic energy as compared to ${\cal E}_0$.  Indeed we
will introduce the important effects of dissipation, by means of the
Caldeira-Leggett formulation, and consider in particular the
interesting case in which dissipation is more important than inertia
\cite{stephen}.  This formally corresponds to the limit
$m{\rightarrow}0$ or ${\gamma}{\rightarrow}0$ at fixed $\eta$.  Note
that in this case the drift velocity in the direction parallel to the
electric field is, classically

\begin{equation} (\dot{\bf r})_{\|}=\frac{eE}{\eta+B^2/\eta}
\label{drift} \end{equation}

\noindent Thus, we will implicitely look at the non-relativistic
limit of a relativistic formulation.  Notice that in the above
formula and in the remainder of the paper we redefine the fields
according to ${\bf E}{\rightarrow}2\pi{\bf E}$ and ${\bf
B}{\rightarrow}2\pi{\bf B}$, unless otherwise stated, so that our
particles have formally charges $e=\pm1$.

The free particle relativistic expression of the energy, Eq.
(\ref{rel}), corresponds to a space-time anisotropic scalar QFT with
Lagrangian

\begin{equation} {\cal
L}_0(\phi)={\partial}_0{\phi}^*{\partial}_0{\phi}-
\frac{1}{\gamma}\nabla{\phi}^*{\cdot}\nabla{\phi} -{\cal
E}_0^2{\phi}^*{\phi} \end{equation}

\noindent Notice that we have introduced a relativistic complex
scalar field $\phi$ describing both vortices and antivortices as its
particle-antiparticle content.  Thus, for a relativistic QFT in which
particle pairs corresponding to vortices and antivortices are
nucleated homogeneously from an external field
$A_{\mu}{\equiv}(V_v,{\bf A}_v)$, the Lagrangian is given by

\begin{equation} {\cal
L}({\phi})=D_0{\phi}^*D_0{\phi}-\frac{1}{\gamma}{\bf
D}{\phi}^*{\cdot}{\bf D}{\phi} -{\cal E}_0^2{\phi}^*{\phi}
\end{equation}

\noindent in which we have introduced the covariant derivative
$D_{\mu}={\partial}_{\mu}-iA_{\mu}$.  Therefore, we see that the
relativistic formulation captures quite naturally the quantum version
of a plausible situation in which the activation energy and the
inertial mass are in general unrelated in the vortex motion through
the supercurrent.  We are therefore in a position to evaluate the
production rate $\Gamma$ of the vortex-antivortex pairs from a
uniform ``electromagnetic'' field $A_{\mu}$ as a function of the
field strengths $E$ and $B$ and of the ratio ${\gamma}=m/{\cal
E}_0$.

\vskip 1.0truecm

\section{ An exact formula for the vortex-antivortex production rate
} \renewcommand{\theequation}{3.\arabic{equation}}
\setcounter{equation}{0}

The calculation that follows is the two-dimensional scalar version of
the well-known Schwinger calculation for the three-dimensional
quantum electrodynamic problem of vacuum decay by production of
electron-positron pairs that are taken to infinity upon creation
\cite{itzzub}.  We evaluate the probability amplitude for the vacuum
decay in time $T$

\begin{equation} Z=\langle 0 | e^{-i\hat{H}T} | 0 \rangle {\equiv}
e^{-iTW_0} \end{equation}

\noindent where $W_0={\cal E}(vac)-i\frac{\Gamma}{2}$ is taken to
give the energy of the vacuum ${\cal E}(vac)$ and its decay rate
$\Gamma$.  With a suitable normalization factor, the probability
amplitude is given by the functional integral over field
configurations

\begin{equation} Z={\cal N}\int {\cal D}{\phi} \exp \left \{-i\int
d^2r dt {\phi}^* \left ( -D_0^2 +\frac{1}{\gamma}{\bf D}^2-{\cal
E}_0^2 \right ) {\phi} \right \} \end{equation}

\noindent which is conveniently evaluated -- formally -- in the
Euclidean metric

\begin{eqnarray} Z&=&\exp \left \{ - Tr \ln \left (
-\frac{1}{\gamma}{\bf D}^2-D_3^2+{\cal E}_0^2 \right ) \right \}
\nonumber \\ &=&\exp (-i T W_0) \label{defin} \end{eqnarray}

\noindent by means of the known identity

\begin{equation} Tr \ln \frac{-D_E^2+{\cal
E}_0^2}{\Lambda^2}=\lim_{{\epsilon}{\rightarrow}0}\int_{\epsilon}^{\infty}
\frac{d\tau}{\tau} Tr \left \{ e^{-(-D_E^2+{\cal E}_0^2)\tau} -
e^{-{\Lambda}^2\tau} \right \} \label{identity} \end{equation}

\noindent in which we have written the square of the covariant
anisotropic Euclidean Laplacian (with $x_3{\equiv}it$):
$D_E^2=D_3^2+\frac{1}{\gamma}(D_1^2+D_2^2)$.

The evaluation of the trace is straightforward with the method of the
Feynman path-integral; for the $(d+1)$-dimensional free particle the
result is, for a system of size $L^d{\times}T$:

\begin{eqnarray} Tr \left \{ e^{-(-{\partial}_E^2)\tau} \right \}
&=&\int dq_0 \langle q_0 | e^{-(-{\partial}_E^2)\tau} | q_0 \rangle
\nonumber \\ &=& \int_{q(0)=q(\tau)=q_0} {\cal D}q(t) \exp \left \{
-\int_0^{\tau} dt \frac{1}{2}m_{\mu} \dot{q}_{\mu}\dot{q}_{\mu}
\right \} \nonumber \\ &=&iT L^d \left ( \frac{1}{4\pi\tau} \right
)^{1/2} \left ( \frac{\gamma}{4\pi\tau} \right )^{d/2} \label{free}
\end{eqnarray}

\noindent where we have employed a fictitious anisotropic mass term
$m_{\mu}$:  $m_1=m_2=\frac{\gamma}{2}$ and $m_3=\frac{1}{2}$.  Hence
$W_0$ is real and, of course, there is no vacuum decay.  In the above
manipulations $q$ is taken to be a $(d+1)$-dimensional coordinate
with anisotropic Euclidean metric; for the particle coupled to a
gauge field case, we have the analogous path-integral construction

\begin{equation} Tr \left \{ e^{-(-D_E^2)\tau} \right \} = \int dq_0
\int_{q(0)=q(\tau)=q_0} {\cal D}q(t) e^{ -\int_0^{\tau} dt {\cal L}_E
} \end{equation}

\noindent where, with the anisotropic fictitious mass terms, the
Euclidean version of the relativistic Lagrangian for the vortex is

\begin{equation} {\cal L}_E=\frac{1}{2} m_{\mu}
\dot{q}_{\mu}\dot{q}_{\mu} - i \dot{q}_{\mu}A_{\mu}(q) \end{equation}

\noindent The uniform-field situation corresponds to
$A_{\mu}(q)=\frac{1}{2}F_{\mu\nu}q_{\nu}$, with the (2+1)-dimensional
field tensor given by

\begin{equation} F_{\mu\nu}= \left ( \begin{array}{ccc} 0 & B & iE_x
\\ -B & 0 & iE_y \\ -iE_x & -iE_y & 0 \end{array} \right )
\end{equation}

\noindent after the analytic continuation to Euclidean time.  The
diagonalization of the operator $(-D_E^2)\tau$ is then achieved via
Fourier transformation in the Euclidean-time variable:

\begin{equation} \int_0^{\tau} dt {\cal L}_E = \frac{\tau}{2\pi}
\sum_n \Omega_{\mu\nu} (\omega_n) q_{\mu}(\omega_n)
q_{\nu}(-\omega_n) \end{equation}

\noindent where

\begin{equation} \Omega_{\mu\nu}(\omega_n)= \left (
\begin{array}{ccc} \frac{\gamma}{4}\omega_n^2 & \frac{\omega_n}{2}B &
\frac{\omega_n}{2}iE_x \\ -\frac{\omega_n}{2}B &
\frac{\gamma}{4}\omega_n^2 & \frac{\omega_n}{2}iE_y \\
-\frac{\omega_n}{2}iE_x & -\frac{\omega_n}{2}iE_y &
\frac{\omega_n^2}{4} \\ \end{array} \right ) \end{equation}

\noindent The path-integral is now readily evaluated, to give

\begin{eqnarray} Tr \left \{ e^{-(-D_E^2)\tau} \right \} &=& iT L^2
{\cal N}(\tau) \prod_{n=1}^{\infty} \left \{ \frac{\tau}{2\pi} \det
\Omega \right \}^{-1} \nonumber \\ &=& iT L^2 {\cal N}(\tau)
\prod_{n=1}^{\infty} \left \{ \left ( \frac{\omega_n^2}{4} \right )^3
{\gamma}^2 \right \}^{-1} \prod_{n=1}^{\infty} \left \{ 1
+\frac{\tau^2}{\pi^2n^2} \left (
-\frac{E^2}{\gamma}+\frac{B^2}{{\gamma}^2} \right ) \right \}^{-1}
\end{eqnarray}

\noindent where the normalization factor ${\cal N}(\tau)$ is adjusted
so as to give the free-particle result, Eq.  (\ref{free}), in the
limit of zero field.  Using the known infinite-product formula
$\prod_{n=1}^{\infty}(1+x^2/\pi^2n^2)={\sinh}(x)/x$, we see that we
get

\begin{equation} Tr \left \{ e^{-(-D_E^2)\tau} \right \} = iT L^2
{\gamma} \left ( \frac{1}{4\pi\tau} \right )^{3/2}
\frac{\tau\sqrt{\frac{E^2}{\gamma}-\frac{B^2}{{\gamma}^2}}} {{\sin}
\left ( \tau\sqrt{\frac{E^2}{\gamma}-\frac{B^2}{{\gamma}^2}} \right )
} \label{trace} \end{equation}

\noindent At this point, the formula for the supercurrent decay rate
becomes, from (\ref{defin}), (\ref{identity}) and (\ref{trace})

\begin{equation} \frac{\Gamma}{2L^2}=Re \int_{\epsilon}^{\infty}
\frac{d\tau}{\tau} i {\gamma} \left ( \frac{1}{4\pi\tau} \right
)^{3/2} e^{-{\cal E}_0^2\tau} \left \{
\frac{\tau\sqrt{\frac{E^2}{\gamma}-\frac{B^2}{{\gamma}^2}}} {{\sin}
\left ( \tau\sqrt{\frac{E^2}{\gamma}-\frac{B^2}{{\gamma}^2}} \right )
} - 1 \right \} \label{rate1} \end{equation}

\noindent where a suitable vacuum subtraction has been inserted (so
that no pair production takes place in the absence of an external
field).  This formula can be evaluated by summing over the residues
at the poles of the $x/{\sin}(x)$ function, namely
$x{\equiv}\tau\sqrt{\frac{E^2}{\gamma}-\frac{B^2}{{\gamma}^2}}=n\pi$,
except for the pole at $x=0$.  To give meaning to the poles one
attributes a small negative imaginary part to $x$, so that $ Im
\frac{1}{{\sin} x} = Im \frac{(-)^n}{x-{\pi}n-i{\epsilon}}=(-)^n
{\pi} {\delta}(x-{\pi}n)$.  The end result is, for the pair
production rate per unit area

\begin{equation} \frac{\Gamma}{L^2}=\frac{\gamma}{4\pi^2} \left (
\sqrt{\frac{E^2}{\gamma}-\frac{B^2}{{\gamma}^2}} \right )^{3/2}
\sum_{n=1}^{\infty} (-)^{n+1} n^{-3/2} \exp \left ( -\frac{{\pi}{\cal
E}_0^2}{\sqrt{\frac{E^2}{\gamma}-\frac{B^2}{{\gamma}^2}}}n \right )
\label{exact} \end{equation}

\noindent The main result in this exact expression, valid for any
ratio $\gamma$, is the presence of a {\it threshold} for the decay of
the supercurrent, $E > E_c{\equiv}B/\sqrt{\gamma}$.  This conclusion
could have been obtained from the classical equation of motion, Eq.
(\ref{classic}).  In fact, from (\ref{classic}) the classical
trajectory, without dissipation i.e.  $\eta=0$, is a cycloid
describing a motion on average orthogonal to ${\bf E}$, that is with
the vortices dragged along with the current ${\bf J}$ (see
Appendix).  The average displacement in the direction parallel to
${\bf E}$ turns out to be $\langle \Delta x \rangle = Em/B^2$, thus
we see that a vortex can nucleate if the energy gained from the
electric field can at least equal the nucleation energy:  $E \langle
\Delta x \rangle {\geq} {\cal E}_0$, which gives the above threshold
condition.  However, even though vortices can nucleate above
threshold, the fact that there is no net transport in the direction
orthogonal to the current in the absence of dissipation implies that
no voltage drop will result in this case, hence no resistence.

In the following we will discuss the interesting and realistic case
in which there is always friction, induced by the quantum dissipation
(for a discussion of a microscopic theory see \cite{niaoth}).  In
this case particles will drift, classically and after a transient
during which a few cycloidal spirals are noted, along a straight line
with drift velocity in the direction of ${\bf E}$ given by Eq.
(\ref{drift}).  Therefore we anticipate the absence of a threshold
for this case, and since a net transport in the direction of ${\bf
E}$ is activated we conclude that a voltage drop will ensue.

As a final comment on the Eq.  (\ref{exact}) we note that, in the
absence of the Magnus field $B$ and at the $n=1$ order -- also
remembering that below the KT transition ${\cal E}_0$ depends on $J$
as in Eq.  (\ref{ren1}) -- the argument of the exponential has the
same leading dependence on the current $J$ as reported in the work of
Ping Ao \cite{ao}.

\vskip 1.0truecm

\section{ Vortex nucleation in the presence of quantum dissipation }
\renewcommand{\theequation}{4.\arabic{equation}}
\setcounter{equation}{0}

Much work in the physics of vortex dynamics assumes the presence of
an amount of dissipation.  At a quantum level, this can be treated
with the formalism introduced by Caldeira and Leggett \cite{calleg}.
This assumes the moving quantum particle to interact with a bath of
quantum harmonic oscillators having an arbitrary distribution of
masses $m_k$ and frequencies $\omega_k$, the only physical
constraints being that the coupling between particle coordinate ${\bf
q}$ and harmonic coordinates ${\bf x}_k$ is linear and the classical
equation of motion reproduces the form (\ref{classic}).  For the
problem at hand, therefore, the Lagrangian describing vortex motion
in the presence of a supercurrent and dissipation is, in Euclidean
time:

\begin{equation} {\cal L}_E^D=
\frac{1}{2}m_{\mu}\dot{q}_{\mu}\dot{q}_{\mu}-\frac{1}{2}i\dot{q}_{\mu}F_{\mu\nu}q_{\nu}
+\sum_k \left \{ \frac{1}{2}m_k (\dot{\bf x}_k^2 + \omega_k^2{\bf
x}_k^2) + c_k{\bf x}_k{\cdot}{\bf q} +
\frac{c_k^2}{2m_k\omega_k^2}{\bf q}^2 \right \} \end{equation}

\noindent where, as is well known, the last term in the oscillators
Lagrangian is added to avoid the lowering of the minimum in the
particle potential by means of the coupling to the external bath, and
where the $c_k$ are constrained by \cite{calleg}

\begin{equation}
\frac{\pi}{2}\sum_k\frac{c_k^2}{m_k\omega_k}{\delta}(\omega-\omega_k){\equiv}J({\omega})
=\eta{\omega} \label{constr} \end{equation}

\noindent $\eta$ being the phenomenological friction coefficient.
The above constraint corresponds to the so-called ohmic case for the
dissipation; more generally, the spectral function can have a
frequency-dependent form (${\omega}_c$ is a cutoff frequency)
$J({\omega})=\eta{\omega}^s\exp(-{\omega}/{\omega}_c)$, where $s>1$
corresponds to the super-ohmic and $s<1$ to the sub-ohmic cases.
Although we deal mainly with the ohmic, $s=1$ case in what follows we
will also extend our results in a rather simple fashion to the
non-ohmic cases in order to show how vortex-pair production is
optimised in the ohmic case.  After Fourier transformation, the ${\bf
x}_k$-modes can be integrated out, for example by means of their
equation of motion in terms of the action ${\cal
S}_E^D=\int_0^{\tau}dt{\cal L}_E^D$

\begin{equation} \frac{{\partial}{\cal S}_E^D}{{\partial}{\bf
x}_k(\omega)}=\frac{\tau}{2\pi} \left ( m_k(\omega^2+\omega_k^2){\bf
x}_k(-\omega)+c_k{\bf q}(-\omega) \right )=0 \end{equation}

\noindent leading to the effective action

\begin{equation} {\cal S}_E^D=\frac{\tau}{2\pi}\sum_n \left \{
q_{\mu}(\omega_n) \Omega_{\mu\nu}(\omega_n)q_{\nu}(-\omega_n) +
\sum_k \frac{c_k^2\omega_n^2}{2m_k\omega_k^2(\omega_n^2+\omega_k^2)}
q_a(\omega_n)q_a(-\omega_n) \right \} \end{equation}

\noindent where the oscillator part contains a summation over the
space-like components $a$=1, 2 only of the particle coordinate.  The
sum over ${\omega}_k$ can be carried out by resorting to the
constraint (\ref{constr}); in fact

\begin{equation} \sum_k
\frac{c_k^2}{2m_k{\omega}_k^2({\omega}_n^2+{\omega}_k^2)}=
\int_0^{+\infty} \frac{d\omega}{{\omega}_n^2+{\omega}^2} \sum_k
\frac{c_k^2}{2m_k{\omega}_k^2}
{\delta}({\omega}-{\omega}_k)=\frac{\eta}{2|{\omega}_n|}
\end{equation}

\noindent We therefore attain the form

\begin{equation} {\cal S}_E^D = \frac{\tau}{2\pi} \sum_n
q_{\mu}(\omega_n) \Omega_{\mu\nu}^D(\omega_n) q_{\nu}(-\omega_n)
\end{equation}

\noindent with $\Omega^D$ modified only in its diagonal space-like
elements by the inclusion of dissipation; for ${\omega}_n>0$

\begin{equation} \Omega_{\mu\nu}^D(\omega_n)= \left (
\begin{array}{ccc}
\frac{\gamma}{4}\omega_n^2+\frac{\eta}{2}{\omega}_n &
\frac{\omega_n}{2}B & \frac{\omega_n}{2}iE_x \\ -\frac{\omega_n}{2}B
& \frac{\gamma}{4}\omega_n^2+\frac{\eta}{2}{\omega}_n &
\frac{\omega_n}{2}iE_y \\ -\frac{\omega_n}{2}iE_x &
-\frac{\omega_n}{2}iE_y & \frac{\omega_n^2}{4} \\ \end{array} \right
) \end{equation}

\noindent Notice that this implies that in Euclidean time dissipation
processes can be represented by means of an effective action.  At
this point, the formula (\ref{rate1}) for the decay rate of the
supercurrent, in the presence of both the uniform external fields and
of dissipation, makes use of the matrix determinant

\begin{equation} \det {\Omega}^D = \frac{{\omega}^2}{4} \left (
\frac{\gamma}{4}{\omega}^2+\frac{\eta}{2}{\omega} \right )^2 -
\frac{{\omega}^2}{4} \left (
\frac{\gamma}{4}{\omega}^2+\frac{\eta}{2}{\omega} \right ) E^2 +
\frac{{\omega}^4}{16}B^2 \end{equation}

\noindent which, adopting a factorization similar to the one employed
without dissipation, leads to

\begin{equation} \frac{\Gamma}{2L^2} = Re \int_{\epsilon}^{\infty}
\frac{d\tau}{\tau} i {\cal N}'(\tau) e^{-{\cal E}_0^2\tau} \left \{
\prod_{n=1}^{\infty} \left [ 1 -
\frac{E^2}{{\eta}{\omega}_n/2+{\gamma}\omega_n^2/4}+\frac{\omega_n^2B^2}
{4({\eta}{\omega}_n/2+{\gamma}\omega_n^2/4)^2} \right ]^{-1} - 1
\right \} \end{equation}

\noindent with the normalization factor ${\cal N}'(\tau)$ given by

\begin{equation} {\cal N}'(\tau) = {\cal N}(\tau)
\prod_{n=1}^{\infty} \left \{ \frac{\omega_n^2}{4} \left (
\frac{\eta}{2}{\omega}_n+\frac{\gamma}{4}{\omega}_n^2 \right )^2
\right \}^{-1} ={\gamma} \left ( \frac{1}{4\pi\tau} \right )^{3/2}
\prod_{n=1}^{\infty} \left ( 1+\frac{2\eta}{{\gamma}{\omega}_n}
\right )^{-1} \end{equation}

\noindent Notice that we have explicitely rewritten the vacuum
subtraction term in such a way that no pair production takes place in
the absence of an external field, $E=B=0$.  In this case a
closed-form expression for $\Gamma$ does not appear possible;
nevertheless the leading contribution to the $\tau$-integral can be
found from the residue of the main, $n=1$ pole of its argument.  No
relevant pole arises from ${\cal N}'(\tau)$, hence the main
contribution is from $\omega=\omega_1={2\pi}/{\tau_1}$ which is the
real positive zero of

\begin{equation} 1 -
\frac{E^2}{{\eta}{\omega}/2+{\gamma}{\omega}^2/4}+\frac{{\omega}^2B^2}
{4({\eta}{\omega}/2+{\gamma}{\omega}^2/4)^2} = 0 \end{equation}

\noindent For given $B$, $\eta$ and $\gamma$, this cubic equation has
a real zero for any infinitesimal electric field $E$, as indeed for
$E{\rightarrow}0$ we get

\begin{equation} {\omega}_1=\frac{2\eta}{\eta^2+B^2} E^2 + \cdots
\label{pole} \end{equation}

\noindent The physical meaning of this is that while in the absence
of dissipation, with
${\omega}_1=2\sqrt{\frac{E^2}{\gamma}-\frac{B^2}{{\gamma}^2}}$, a
real root appears only for $E>B/\sqrt{\gamma}$, now any infinitesimal
$E$, therefore any infinitesimal supercurrent, gives rise to vortex
pair production and decay.  The main consequence of (ohmic)
dissipation is therefore the elimination of the threshold for a
non-vanishing $\Gamma$, as was also noted by , e.g., Ao and Thouless
\cite{thouless}.  Noticing that $\omega_1$ is independent of
$\gamma$, we work out the final formula in the
``dissipaton-dominated'' case, that is in the limit
${\gamma}{\rightarrow}0$ also considered by Stephen \cite{stephen},
although a less instructive analytic result could be obtained for
arbitrary $\gamma$ to leading order in $E^2$.  The formula for
$\Gamma$ is first of all rewritten, more conveniently, as

\begin{eqnarray} \frac{\Gamma}{2L^2} &=& Re \int_{\epsilon}^{\infty}
\frac{d\tau}{\tau} i \gamma \left ( \frac{1}{4\pi\tau} \right )^{3/2}
\prod_{n=1}^{\infty} \left \{ 1+4\frac{
B^2+\eta^2+{\gamma}(\eta{\omega}_n-E^2) }{ {\gamma}^2{\omega}_n^2 }
\right \}^{-1} \nonumber \\ &\cdot& e^{-{\cal E}_0^2\tau} \left [
\prod_{n=1}^{\infty} \left \{ 1 - \frac{ 8{\eta}E^2 } {
{\gamma}^2{\omega}_n^3 + 4{\omega}_n \left (
B^2+\eta^2+{\gamma}(\eta{\omega}_n-E^2) \right ) } \right \}^{-1} -1
\right ] \end{eqnarray}

\noindent which for ${\gamma}{\rightarrow}0$ simplifies to

\begin{equation} \lim \frac{\Gamma}{2L^2}=Re \int_{\epsilon}^{\infty}
\frac{d\tau}{\tau} i {\cal N}''(\tau) e^{-{\cal E}_0^2\tau} \left \{
\prod_{n=1}^{\infty} \left ( 1-\frac{ 2{\eta}E^2 } {
{\omega}_n(B^2+\eta^2) } \right )^{-1} - 1 \right \} \end{equation}

\noindent with the following asymptotic form of the normalization
factor

\begin{eqnarray} {\cal N}''(\tau)&=&{\gamma} \left (
\frac{1}{4\pi\tau} \right )^{3/2} \prod_{n=1}^{\infty} \left (
1+4\frac{B^2+\eta^2}{{\gamma}^2{\omega}_n^2} \right )^{-1} \nonumber
\\ && {\rightarrow} {\gamma} \left ( \frac{1}{4\pi\tau} \right
)^{3/2} 2 \frac{\tau}{\gamma} \sqrt{B^2+\eta^2} e^{
-\frac{\tau}{\gamma}\sqrt{B^2+\eta^2} } \end{eqnarray}

\noindent in which the limit form $x/\sinh(x){\rightarrow}2xe^{-x}$
has been used.  The leading pole has the precise form (\ref{pole}),
but for arbitrary $E$, and evaluating the residue at this pole the
formula for $\Gamma$ reads, with ${\cal E}_{0R}^2={\cal
E}_0^2+\frac{1}{\gamma}\sqrt{B^2+\eta^2}$

\begin{equation} \lim \frac{\Gamma}{2L^2}=2\pi \tau_1 e^{-{\cal
E}_{0R}^2\tau_1} \left ( \frac{1}{4\pi\tau_1} \right )^{3/2}
\sqrt{B^2+\eta^2} \prod_{n=2}^{\infty} \left ( 1-\frac{1}{n} \right
)^{-1} \label{gammatau1} \end{equation}

\noindent The last infinite product is clearly divergent unless a
frequency cutoff is introduced, $\omega_n{\leq}\omega_c$ where
$\omega_c$ is a large frequency above which the effects of
dissipation can be neglected \cite{calleg}.  This implies the
existence of a cutoff integer $n^*=[\omega_c/\omega_1]$ and the last
factor will thus be a cutoff-dependent dimensionless number
$C(n^*)=C^*$.  For small $E^2$ we see that
$n^*=\frac{2{\eta}}{{\eta}^2+B^2}\frac{{\omega}_c}{E^2}$ is large and
in this case we can estimate the asymptotic behaviour of $C^*$ by

\begin{equation} C^*=\prod_{n=2}^{n^*} \left ( 1-\frac{1}{n} \right
)^{-1} {\rightarrow} \exp -n^* \int_{2/n^*}^1 dx \ln \left (
1-\frac{1}{n^* x} \right ) {\simeq} \frac{e}{4} n^* =
\frac{e{\eta}}{{\eta}^2+B^2} \frac{{\omega}_c}{2E^2} \label{limprod}
\end{equation}

\noindent If, on the other hand, $E^2$ is very large and formally
$n^*$ shrinks to zero, it would mean that our situation is such that
the Caldeira-Leggett dissipative regime breaks down.  We assume here
to be always in a range of $E$ compatible with the Caldeira-Leggett
regime, thus $n^*$ can be either large (corresponding to $C^*$ given
by (\ref{limprod})) or finite (corresponding to $C^*$ of order
unity).  The final asymptotic form for the pair production rate when
${\gamma}{\rightarrow}0$ will be, therefore, substituting
$\tau_1=2{\pi}/{\omega}_1$ from Eq.  (\ref{pole}) in Eq.
(\ref{gammatau1}),

\begin{equation} \lim \frac{\Gamma}{L^2} {\simeq} \frac{1}{2\pi} C^*
\sqrt{\eta} E \exp \left \{ -{\cal
E}_{0R}^2\frac{\pi(B^2+\eta^2)}{{\eta}E^2} \right \} \label{fric}
\end{equation}

\noindent where we have defined the ``renormalized'' activation
energy

\begin{equation} {\cal E}_{0R}^2={\cal E}_0^2 + \frac{1}{\gamma}
\sqrt{B^2+\eta^2} \label{massren} \end{equation}

\noindent This has a physical interpretation connected with the
energy $B/2m$ of the lowest Landau level which becomes infinite in
the limit of zero inertial mass and contributes a diverging
zero-point energy to the total activation energy.  Indeed, for
${\eta}=0$, ${\cal E}_{0R}{\simeq}{\cal E}_0+B^2/2m$.  If an infinite
barrier for placing a vortex in the lowest Landau level does really
develop, then vortex pair production will cease and there will be no
supercurrent decay.  We therefore generically assume in the following
that ${\cal E}_{0R}$ is ultimately just a phenomenological
finite-value parameter of the theory.

At this point we can generalise the results obtained above for the
ohmic, $s=1$, case to the non-ohmic, $s\not=1$, situations (for a
recent discussion on the relevance of non-ohmic cases, see
\cite{guinea}).  Indeed, Eq.  (\ref{pole}) holds good if
$J({\omega})=\eta{\omega}^s$ provided we replace $\eta$ with a
frequency-dependent friction coefficient
$\eta({\omega})=\eta{\omega}^p$, with $p=s-1$.  Then we have the
following equation for the main pole ${\omega}_1=2\pi/{\tau}_1$

\begin{equation}
{\omega}_1=\frac{2\eta{\omega}_1^p}{\eta^2{\omega}_1^{2p}+B^2}E^2+{\cdots}
\label{nonohmic} \end{equation}

\noindent which we can qualitatively handle in the limit
$E{\rightarrow}0$.  First, we observe from Eq.  (\ref{nonohmic}) that
the role of $B$ is enhanced in the super-ohmic case, while it is
suppressed if dissipation is sub-ohmic, in agreement with
\cite{thouless}.  Then, both in the super-ohmic $p>0$ and in the
sub-ohmic $p<0$ cases, we get that ${\omega}_1{\simeq}E^{2/(1-|p|)}$
so that in the relevant formula for $\Gamma$, e.g.  Eq.
(\ref{fric}), we must substitute $E$ with $E^{1/(1-|p|)}$.  For small
$E$ this leads to a suppression of the vortex production rate as soon
as $p\not=0$, compared to the ohmic case which therefore represents
the situation with the maximum production rate.

To conclude this Section, we point out that the argument of the
exponential appearing in Eq.  (\ref{fric}) has a direct physical
interpretation.  Apart from constants, it is the Action
(=Energy${\times}$Time) required for the nucleation of a vortex,
since Energy=${\cal E}_{0R}$ and Time=${\ell}_N/v_N=\frac{{\cal
E}_{0R}}{E}{\cdot}\frac{\eta(B)}{E}$.  Indeed, ${\ell}_N=\frac{{\cal
E}_{0R}}{E}$ is the distance involved for the nucleation in a field
$E$ whereas $v_N=\frac{E}{\eta(B)}$ is the drift velocity (if we
indicate $\eta(B)=(\eta^2+B^2)/\eta$).

\vskip 1.0truecm

\section{ Vortex nucleation in the presence of a pinning potential }
\renewcommand{\theequation}{5.\arabic{equation}}
\setcounter{equation}{0}

Most superconducting materials in which vortex dynamics plays a role
contain a finite density of defects of some sort.  Many discussions
in the literature deal with the tunneling effect of vortices trapped
by a potential barrier on the resistence of a superconductor.  A
finite height and width of the pinning barrier in the direction of
the ``electric field'' ${\bf E}$, that is perpendicular to the
current ${\bf J}$, ultimately give rise to a finite additional term
which renormalises ${\cal E}_{0R}$.  Pinning barriers in the
direction of the current have instead a different physical effect, in
particular in the dissipation-dominated case.  In our own
relativistic formulation we can account for an additional effect due
to pinning if we consider the ``extreme'' case of the unbounded
harmonic-well potential

\begin{equation} U({\bf q})=\frac{1}{2}k_xq_x^2+\frac{1}{2}k_yq_y^2
\label{well} \end{equation}

\noindent which we add to our quantum field Lagrangian through the
term $U({\bf r}){\phi}^*{\phi}$.  Strictly-speaking, in our
relativistic treatment this amounts to a confinement force that,
while linear in ${\bf q}$ at small distances, at large distances
remains constant and thus simulates the presence of a uniform
``confinement pressure'' on the vortices.  Thus, in this ``extreme''
case a threshold for the pair production may reappear.  In what
follows, we first treat the case of the unbounded potential in the
absence of dissipation, and then discuss the more realistic case of a
finite potential barrier.  Technically, with the form (\ref{well}) we
can formally repeat the calculation of the previous Sections, with a
quadratic form in the path integral defined by the matrix

\begin{equation} \tilde{\Omega}_{\mu\nu}(\omega_n)= \left (
\begin{array}{ccc}
\frac{\gamma}{4}\omega_n^2+\frac{\eta}{2}{\omega}_n+\frac{1}{2}k_x &
\frac{\omega_n}{2}B & \frac{\omega_n}{2}iE_x \\ -\frac{\omega_n}{2}B
& \frac{\gamma}{4}\omega_n^2+\frac{\eta}{2}{\omega}_n+\frac{1}{2}k_y
& \frac{\omega_n}{2}iE_y \\ -\frac{\omega_n}{2}iE_x &
-\frac{\omega_n}{2}iE_y & \frac{\omega_n^2}{4} \\ \end{array} \right
) \end{equation}

Having the limit ${\gamma}{\rightarrow}0$ always in mind, we begin
the discussion of this extreme case of pinning by considering the
absence of dissipation, $\eta=0$.  Then, with a suitable
factorization for $\det \tilde{\Omega}$ we can write, for the pair
production rate and in the limit ${\gamma}{\rightarrow}0$

\begin{equation} \lim \frac{\Gamma}{2L^2}=Re \int_{\epsilon}^{\infty}
\frac{d\tau}{\tau} i {\cal N}'(\tau) e^{-{\cal E}_0^2\tau} \left \{
\prod_{n=1}^{\infty} \left ( 1+\frac{k_xk_y -
2(k_xE_y^2+k_yE_x^2)}{{\omega}_n^2B^2} \right )^{-1} - {\rm const.}
\right \} \end{equation}

\noindent The normalization factor has again a singular limit form

\begin{eqnarray} {\cal N}'(\tau)&=&{\gamma} \left (
\frac{1}{4\pi\tau} \right )^{3/2} \prod_{n=1}^{\infty} \left \{
1+\frac{4}{{\gamma}^2{\omega}_n^2} \left (
{\gamma}\frac{k_x+k_y}{2}-{\gamma}E^2+B^2 \right ) \right \}^{-1}
\nonumber \\ &&{\rightarrow} \left ( \frac{1}{4\pi\tau} \right
)^{3/2} 2{\tau}Be^{-\frac{\tau}{\gamma}B} \end{eqnarray}

\noindent in which we see that the very same quantum zero-point
energy fluctuations in the Landau levels give rise to a divergent
renormalization in the rest mass for ${\gamma}{\rightarrow}0$:
${\cal E}_0^2{\rightarrow}{\cal E}_{0R}^2={\cal E}_0^2+B/{\gamma}$.
The remainder of the integral in $\tau$ can be evaluated as in
Section III in this limit, and we end up with the exact formula

\begin{eqnarray} \lim \frac{\Gamma}{L^2}&=&\frac{\sqrt{B}}{2\pi}
\left \{ \frac{1}{2} \left ( k_xE_y^2 +k_yE_x^2 - k_xk_y/2 \right )
\right \}^{1/4} \nonumber \\ &&{\cdot} \sum_{n=1}^{\infty} (-1)^{n+1}
n^{-1/2} \exp \left \{ -\frac{{\pi}B{\cal E}_{0R}^2}
{\sqrt{\frac{1}{2}(k_xE_y^2+k_yE_x^2-k_xk_y/2)}} n \right \}
\end{eqnarray}

\noindent in which we see that $\Gamma$ is zero unless the condition
$2E_x^2/k_x +2E_y^2/k_y>1$ is satisfied, that is no production of
vortex-antivortex pairs takes place inside an ellipse in $(E_x,E_y)$
having half-axes $\sqrt{k_x/2}$ and $\sqrt{k_y/2}$.  The physical
interpretation of this phenomenon is that the threshold in the
minimum value of the electric field for a non-vanishing pair
production is now restored by confinement.  Indeed, while particles
get trapped in stable Landau levels for a vanishing inertial mass,
under the action of a pinning potential the trapping role of the
magnetic field is restricted to a renormalization of the rest mass.
Transitions out of these levels can occur for strong enough electric
fields since the situation becomes now a quasi-static equilibrium
between the confinement and the electrostatic forces, the dynamics
being separated away with the magnetic field term.  Notice, however,
that for $E_y=0$ and $k_x=0$ (current and confinement both in the
$y$-direction) there is a non-vanishing rate $\Gamma$ for any
$E_x{\not=}0$, hence there is no threshold.  Nevertheless, no net
transport takes place in the direction of the electric field and the
supercurrent does not decay if dissipation is absent (see Appendix).

We now consider the case in which ohmic dissipation is present in a
more realistic setting, by specializing to $E_y=k_x=0$.  In fact, one
can take into account a finite-width pinning barrier in the
$x$-direction by a renormalization of ${\cal E}_0$, but a new
qualitative feature appears if pinning is in the direction of the
current, ${\bf J}=-{\times}{\bf E}$.  We adopt the following
factorization for the matrix determinant:

\begin{eqnarray} \det \tilde{\Omega}^D&=&\frac{{\omega}^2}{4} \left (
\frac{\gamma}{4} {\omega}^2 \right )^2 \left \{ 1 + \frac{
4{\eta}^2+4B^2+{\gamma}(4{\eta}{\omega}+2k_y-4E^2) }
{{\gamma}^2{\omega}^2} \right \} \nonumber \\ &{\times}& \left \{ 1 +
\frac{ \left ( 4k_y-8E^2 \right ){\eta}{\omega}^{-1} - 8k_yE^2
{\omega}^{-2} } {
{\gamma}^2{\omega}^2+4{\eta}^2+4B^2+{\gamma}(4{\eta}{\omega}+2k_y-4E^2)
} \right \} \end{eqnarray}

\noindent since for ${\gamma}{\rightarrow}0$ this becomes

\begin{eqnarray} \lim \det \tilde{\Omega}^D&=&\frac{{\omega}^2}{4}
\left ( \frac{\gamma}{4} {\omega}^2 \right )^2 \left \{ 1 +
4\frac{B^2+{\eta}^2}{{\gamma}^2{\omega}^2} \right \} \nonumber
\\ &{\times}& \left \{ 1 + \frac{1}{B^2+{\eta}^2} \left [
\frac{k_y-2E^2}{\omega} {\eta} + \frac{-2k_yE^2}{{\omega}^2} \right ]
\right \} \end{eqnarray}

\noindent Here we see that the prefactor is of the type already seen
for dissipation, hence it gives rise to a nucleation energy
renormalization of the type (\ref{massren}).  Denoting

\begin{eqnarray} \tilde{B}^2&=&B^2+{\eta}^2 \nonumber \\
\tilde{E}^2&=&E^2-\frac{1}{2}k_y \nonumber
\\ \tilde{K}^2&=&\frac{1}{2}k_yE^2 \end{eqnarray}

\noindent we see that the dominant pole in the integral formula for
$\Gamma$ is the solution of
$\tilde{B}^2{\omega}_1^2-2{\eta}\tilde{E}^2{\omega}_1-4\tilde{K}^2=0$.
The only relevant principal pole is then for

\begin{equation} {\omega}_1=\frac{2\pi}{\tau_1}=\left (
\eta\tilde{E}^2+\sqrt{ \eta^2\tilde{E}^4 +4\tilde{B}^2\tilde{K}^2 }
\right ) / \tilde{B}^2 \label{ppole} \end{equation}

\noindent for which the residue is readily calculated and yields the
main contribution to the vacuum decay rate

\begin{equation} \lim \frac{\Gamma}{L^2}{\simeq}2{\pi}e^{-{\cal
E}_{0R}^2{\tau}_1} \left ( \frac{1}{4\pi{\tau}_1} \right )^{3/2} C^*
\sqrt{B^2+\eta^2} \end{equation}

\noindent where $C^*$ is expressed by an infinite product, cutoff at
$n^*={\omega}_c/{\omega}_1$, similarly to what seen in Section IV.
The formula for the pole, Eq.  (\ref{ppole}), reveals its physical
contents in the limit in which the $E^2{\ll}k_y$.  Expanding the
square-root to lowest order, we always obtain a pole for small $E$

\begin{equation} {\omega}_1=2\frac{E^2}{\eta} \label{pinn}
\end{equation}

\noindent which is of the same form as Eq.  (\ref{pole}) for the
purely dissipative case, except that -- due to pinning -- the
renormalization effect induced by the Magnus field on the friction
coefficient has disappeared.  This result is in agreement with the
analysis based on the classical equations presented in the Appendix,
which, furthermore, shows that Eq.  (\ref{pinn}) holds (for small
$E$) also for a finite-height and finite-width pinning barrier.  In
the general case in which the pinning centres will be distributed
with some density, this will correspond to an effective friction
coefficient that will present a dependence on $B$ intermediate
between the two cases:  $\eta<\eta_{eff}<(\eta^2+B^2)/\eta$.

\vskip 1.0truecm

\section{ An Approach to Thermal Effects and Conclusions }
\renewcommand{\theequation}{6.\arabic{equation}}
\setcounter{equation}{0}

To complete the picture one would need to evaluate the vortex
nucleation rate under the effect of both the external current $J$
(resulting in the ``electric'' field $E$) and the temperature $T$ in
order to estimate the voltage drop.  This appears to be a rather
difficult task, since it would involve a theoretical description of
how the thermal distribution of vortices is obtained from a state
with no vortices, a process of non-equilibrium thermodynamics.

Here we will take a less ambitious view and try to obtain a
phenomenological description of the process of vortex pair production
in the presence of thermal fluctuations.  By a detailed balance
argument, to be developed below, one finds that the density of free
vortices $\rho_f$ is proportional to the square root of the
production rate, $\rho_f{\sim}\sqrt{\Gamma}$.  Considering the
relevant case of the dissipation-dominated dynamics described in
Section III, we tentatively propose to simulate the effect of
temperature on the vortex production by introducing thermal currents
$J(T)$ (induced for instance by phonons, or otherwise) and using the
result of Eq.  (\ref{fric}), with $E^2=(2\pi J(T))^2$, namely
${\Gamma}{\sim}\exp -{\cal E}_{0R}^2\eta_{eff}/4\pi J^2(T)$.  The
requirement of obtaining a Boltzmann distribution for $\rho_f$ fixes
$J^2(T)=\frac{1}{8\pi}\eta_{eff}{\cal E}_{0R}T$ (we take $k_B=1$ for
Boltzmann's constant in what follows).  In the general case, where we
have both temperature $T$ and external current $J$, we will use Eq.
(\ref{fric}) with $E^2=(2\pi J_{tot})^2$ by adding incoherently the
two contributions:  $J_{tot}^2=J(T)^2+J^2$.  We can thus rewrite, in
the general case

\begin{equation} \frac{\Gamma}{L_xL_y}{\simeq}C^*\sqrt{\eta}J_{tot}
\exp -2{\cal E}_{0R}/(T+{\Delta}T(J)) \end{equation}

\noindent where $L_{x,y}$ are the linear sizes of the sample; also

\begin{equation} {\Delta}T(J)=\frac{8\pi J^2}{\eta_{eff}{\cal
E}_{0R}}~{\sim}~t_n(J)^{-1} \end{equation}

\noindent and we have indicated that ${\Delta}T(J)$ is proportional
to the inverse of the nucleation time for a vortex (see the
discussion at the end of Section IV).  Let us recall that below the
KT transition ${\cal E}_{0R}$ depends on $J$ (see Eq.  (\ref{ren1}))
and in this dependence we include just the external coherent current
as $<J_{tot}>=J$.  Therefore, vortices are not produced for zero
current and temperatures below the KT transition.

We can then compute the free vortex number density $\rho_f$ by a
detailed balance argument:  the production rate must equal the
annihilation rate.  Neglecting the rate at which the vortices
disappear from the sample due to the drift described by Eq.
(\ref{drift}), thus assuming macroscopic dimensions for the sample,
the annihilation rate per unit area will be due to two mechanisms:
\par\noindent 1) A possible vortex-antivortex annihilation, and this
rate will be proportional to the product of an annihilation length
$\sigma$ (the 2-D analogue of the familiar 3-D cross section), times
the density of the free vortices $\rho_f$, times the incident flux
$\rho_f{\cdot} 2<v>$, where the average drift velocity will be given,
according to (\ref{drift}), by $\langle v
\rangle=2{\pi}J_{tot}/\eta_{eff}$.  Hence:

\begin{equation} \frac{{\Gamma}_{Af}}{L_xL_y} {\simeq} 2 \rho_f^2
\langle v \rangle \sigma \end{equation}

\noindent 2) A possible annihilation of vortices with pinned
antivortices, with a rate

\begin{equation} \frac{{\Gamma}_{Ap}}{L_xL_y} {\simeq} \rho_f\rho_p
\langle v \rangle \sigma \end{equation}

\noindent where $\rho_p$ is the density of pinned vortices.  In turn,
at equilibrium we will have $\rho_f{\Gamma}_p=\rho_p{\Gamma}_u$,
where ${\Gamma}_p$ and ${\Gamma}_u$ are the rates for pinning and
unpinning, respectively.  In conclusion, the detailed balance will be

\begin{equation} \frac{\Gamma}{L_xL_y}=\rho_f^2 \sigma
\frac{2{\pi}J_{tot}}{\eta_{eff}} \left ( 2 + {\Gamma}_p/{\Gamma}_u
\right ) \end{equation}

\noindent which gives

\begin{equation} \rho_f=\left [ \frac{ C^* \eta^\frac{1}{2}\eta_{eff}
} {2\pi \sigma (2+{\Gamma}_p/{\Gamma}_u) } \right ]^\frac{1}{2} \exp
- \frac{ {\cal E}_{0R} } { T+{\Delta}T(J) } \end{equation}

\noindent Thus, the vortex current ${\bf J}_v$ in the $x$-direction
orthogonal to the supercurrent density ${\bf J}$ will be
$J_v=2\pi\rho_fJ/\eta_{eff}$, yielding the potential drop
${\Delta}V=L_xJ_v$ in the $y$-direction parallel to ${\bf J}$.
Thus:

\begin{equation} {\Delta}V=2\pi L_x J \left [ \frac{ C^*
\eta^\frac{1}{2} }{\eta_{eff} \sigma(2+{\Gamma}_p/{\Gamma}_u) }
\right ]^\frac{1}{2} \exp - \frac{ {\cal E}_{0R} }{ T+{\Delta}T(J) }
\end{equation}

\noindent For small external currents such that ${\Delta}T(J){\ll}T$
one gets, above the KT transition, in terms of the current $I=L_yJ$

\begin{equation} {\Delta}V{\sim}R(T)I+Q(T)I^3+{\cdots} \end{equation}

\noindent where

\begin{eqnarray} R(T)&=&2\pi \frac{L_x}{L_y} \left [ \frac{ C^*
\eta^\frac{1}{2} }{ \eta_{eff}\sigma(2+{\Gamma}_p/{\Gamma}_u) }
\right ]^\frac{1}{2} e^{-{\cal E}_{0R}/T} \nonumber
\\ Q(T)&=&\frac{R(T)}{L_y^2} \frac{8\pi}{\eta_{eff}T^2}
\label{resist} \end{eqnarray}

\noindent It is to be stressed at this point that the above
considerations have led us to a rather predictable result:  there is
no resistance in the sample if dissipation is absent.  As we have
previously pointed out, this qualitative conclusion could have been
obtained from the classical equations of motion, Eq.s (\ref{classic})
and (\ref{drift}) (see Appendix).  We recall that, as discussed at
the end of Section V and in the Appendix, in the general case of a
distribution of pinning barriers one will have
$\eta<\eta_{eff}<(\eta^2+4\pi^2\rho_s^2)/\eta$ (remembering that
$B=2\pi\rho_s$).

We finally point out that, for temperatures below the KT transition,
${\cal E}_{0R}$ depends on $J$, as in Eq.  (\ref{ren1}).  Thus, for
low $J$ such that ${\Delta}T(J){\ll}T$, we recover the standard
result that ${\Delta}V{\sim}I^{1+a}$, where $a{\rightarrow}2$ for
$T{\rightarrow}T_{KT}$ \cite{shenoy}.  Our formula includes dynamical
effects due to current-induced nucleation.  In particular, by
expanding for low $J$, we can repeat the above calculations to arrive
at the following dependence on $J$

\begin{equation} {\Delta}V{\sim}c_1I^{1+a}+c_2I^{3+a} \left ( \ln
\frac{m_0J}{\rho_s a} \right )^2 + {\cdots} \end{equation}

\noindent where $c_1$ and $c_2$ have expressions similar to those
reported in (\ref{resist}).

To conclude, we have presented a plausible physical scenario for the
applicability of our exact formalism giving the decay rate of a
superconducting current and thus the resistence, or voltage drop,
observable in a real sample.  Thermal effects have been accounted for
phenomenologically through thermally-induced currents, and pinning
forces by means of a harmonic potential.  We find that in some
extremal cases there is a threshold for the production of
vortex-antivortex pairs, in that these are trapped in Landau or
harmonic-oscillator levels, but that the threshold is eliminated, if
extremal pinning is absent, by the addition of any infinitesimal
amount of dissipation.  In the case of finite pinning, or even
infinite-pinning barrier in the direction of the current, the
threshold disappears.  Pinning barriers in the direction of the
current suppress the effect of the Magnus force (confined to a
renormalization of the vortex nucleation energy).  The limit of a
vanishing inertial mass has been shown to lead to considerable
simplification in the algebra, as well as to new physical effects
like the divergent renormalization of the vortex nucleation energy,
resulting in the inhibition of vortex production for all the cases
considered when the inertial mass is strictly zero.  Further work is
needed to account for the thermal effects in the nucleation of
vortices within a fully microscopic theoretical approach.

\newpage

\vskip 1.0truecm

\section*{ Appendix }
\renewcommand{\theequation}{A.\arabic{equation}}
\setcounter{equation}{0}

Here we summarize some relevant results from the solution of the
classical equations of motion for the vortex dynamics, Eq.
(\ref{classic}) (for charges with $e^2=1$).  We begin by recalling
that in the presence of the fields ${\bf E}=E\hat{\bf x}$ and ${\bf
B}=B{\hat{\bf z}}$ the solution corresponding to the initial
conditions $x(0)=y(0)=0$ and $\dot{x}(0)=\dot{y}(0)=0$ is of the form

\begin{eqnarray} x(t)&=&\frac{mE}{eB^2}(1-\cos{\omega}_ct) \nonumber
\\ y(t)&=&-\frac{mE}{eB^2}\sin{\omega}_ct+\frac{E}{B}t \end{eqnarray}

\noindent which represents a cycloid with frequency
${\omega}_c=eB/m$.  This shows that the average displacement along
the electric field's $x$-direction is fixed at $\langle x(t)
\rangle=x_0=mE/eB^2$ so that there is no transport in the direction
orthogonal to the current, hence no voltage drop.  Also, we require
the field to be such that $Ex_0{\geq}{\cal E}_0$ to extract the
particle from its classical trajectory, in agreement with our
conclusion on the existence of a threshold in Section III.  If on the
contrary friction is switched on, the particle will drift after a
transient along a straight line with a drift velocity in the
$x$-direction given precisely by Eq.  (\ref{drift}); in this case
transport is activated and a voltage drop ensues.  Indeed, with the
same initial conditions as above, the motion in the presence of
friction is represented by

\begin{eqnarray} x(t)&=&-\frac{meE}{\eta^2+B^2} \left \{
e^{-(\eta/m)t} \cos({\omega}_ct-2\alpha ) -\cos2\alpha \right
\}+\frac{{\eta}eE}{\eta^2+B^2}t \nonumber
\\ y(t)&=&-\frac{meE}{\eta^2+B^2}\left \{ e^{-(\eta/m)t}
\sin({\omega}_ct-2\alpha )+\sin2\alpha \right \}
+\frac{BE}{\eta^2+B^2}t \end{eqnarray}

\noindent where $\tan\alpha =\eta /B$, showing that (after a
transient damped-cycloidal motion that can be neglected for
$t{\geq}m/\eta$) we have an average $x$-component of the velocity
$\langle \dot{x}(t) \rangle=eE/\eta(B)$ with
$\eta(B)=(\eta^2+B^2)/\eta$ as in Eq.  (\ref{drift}).  Also, there is
no threshold value of $E$ for the vortex-antivortex nucleation.

Next we consider what might be, classically, the main effect of
pinning on the vortex motion.  We introduce a pinning potential
approximately described by
$U(x,y)=\frac{1}{2}k_x(x-x_p)^2+\frac{1}{2}k_y(y-y_p)^2$ for
$|x-x_p|<{\ell}_x$ and $|y-y_p|<{\ell}_y$ and zero otherwise, where
$(x_p,y_p)$ are the coordinates of the pinning centre.  We imagine
that there will be many pinning centres in the sample, but discuss
the local dynamics of a vortex around one of the impurity sites.  Our
particles, i.e.  the vortices, will feel a potential barrier around
$(x_p,y_p)$.  Like in Section V, we assume that the potential barrier
in the direction of the electric field ${\bf E}=E\hat{\bf x}$ will
have the main effect of increasing the activation energy ${\cal
E}_{0R}$, thus renormalising it by way of the addition of a further
term.  We can thus treat this part of the problem
phenomenologically.  The barrier in the $y$-direction has a different
implication for the dynamics.  Consider first the frictionless case,
$\eta=0$.  The general solution of the classical equation, including
the force due to $U(y)$, is

\begin{eqnarray}
x(t)-x_p&=&-\frac{{\omega}_c}{\omega}A\cos{\omega}\tau
+\frac{1}{2}\frac{k_yeE}{mk_y+B^2}\tau^2 +\frac{k_y}{eB}y_0\tau +x_0
\nonumber \\ y(t)-y_p&=&-A\sin{\omega}\tau + \frac{EB}{mk_y+B^2}\tau
+y_0 \end{eqnarray}

\noindent where $\tau =t-t_0$ and
${\omega}=\sqrt{{\omega}_c^2+k_y/m}$.  We see that, after averaging
over the oscillations and with general initial conditions, $\langle
y(t)-y_p \rangle=\frac{EB}{mk_y+B^2}\tau +y_0$ and $\langle x(t)-x_p
\rangle=x_0+\frac{k_y}{eB}y_0\tau +
\frac{1}{2}\frac{k_yeE}{mk_y+B^2}\tau^2$, so that under the effect of
pinning in the $y$-direction $\langle x(t)-x_p \rangle$ can reach,
after a suitable time, a large enough value and allow the vortex to
gain nucleation energy from the electric field.  Depending on the
values of the parameters this can occur within the pinning potential
width ${\ell}_y$ , in which case the pinning in the $y$-direction
removes the threshold for vortex nucleation which occurred without
friction and pinning.  However, the classical equations indicate that
even in the present case there is no net transport in the
$x$-direction if there is no friction.  In fact, the previous
solution holds up to $|\langle y(t)-y_p \rangle|{\sim}{\ell}_y$,
beyond which either there is no longer any pinning, and thus the
motion will fall back on the familiar cycloid (resulting in no
transport), or else the vortex will feel the effect of another
pinning centre at $(x_p',y_p')$.  In the latter case we will have, at
the initial time $t_0$ of the new piece of the trajectory, $\langle
y(t_0)-y_p' \rangle=-{\ell}_y'$.  Therefore, at the later time $t_1$
for which $\langle y(t_1)-y_p' \rangle={\ell}_y'$ (that is as the
vortex moves through the $y$-range of the new pinning centre from
$y=y_p'-{\ell}_y'$ to $y=y_p'+{\ell}_p'$), we find that $\langle
x(t_1)-x_p' \rangle = \langle x(t_0)-x_p' \rangle$.  The conclusion
is that there has been no net transport in the $x$-direction.  The
situation is completely different if there is friction, $\eta\not=0$,
as can be easily seen by solving the classical equations by formally
taking $m\ddot{x}=0$ and $m\ddot{y}=0$ for long times and adding the
pinning force due to $U(y)$.  The solution is that, for long times
($t>m/\eta$), $\langle \dot{x}(t) \rangle = eE/\eta$ whilst $\langle
y(t)-y_p' \rangle = BE/k_y\eta$.  Thus, there is always the
possibility of nucleation and transport.  Comparing with the case
without pinning, we see that the drift velocity in the $x$-direction
is now determined by the actual friction coefficient
$\eta<\eta(B)=(\eta^2+B^2)/\eta$, whilst in $y$ one has simply a
displacement (thus, for low $E$ at least, $y(t)$ will remain in the
range where $U(y)\not=0$).  In the general case, where there can be
pinning centers distributed with some density, the effective friction
coefficient $\eta_{eff}$ will be somehow in between the extreme
cases, $\eta<\eta_{eff}<(\eta^2+B^2)/\eta$.

\begin{center} {\bf Acknowledgements } \end{center}

One of us (G.J.)  is grateful to the International School for
Advanced Studies in Trieste, where part of his work was carried out,
for hospitality and use of its facilities.


\newpage

\begin{thebibliography}{99} \bibitem{geilo} For a recent survey of
this field, see the articles in:  {\em Phase Transitions and
Relaxation in Systems with Competing Energy Scales}, T.  Riste and
D.  Sherrington (Ed.s) (Kluwer, Dordrecht 1993).  \bibitem{blatter}
G.  Blatter, V.B.  Geshkenbain and V.M.  Vinokur, {\em Phys.  Rev.
Lett.} {\bf 66} 3297 (1991); G.  Blatter and V.B.  Geshkenbein, {\em
Phys.  Rev.}  {\bf B47} 2725 (1993).  \bibitem{ivlev} B.I.  Ivlev,
Yu.N.  Ovchinnikov and R.S.  Thompson, {\em Phys.  Rev.}  {\bf B44}
7023 (1991).  \bibitem{thouless} Ping Ao and D.J.  Thouless, {\em
Phys.  Rev.  Lett.}  {\bf 72}, 132 (1994).  \bibitem{stephen} M.J.
Stephen, {\em Phys.  Rev.  Lett.}  {\bf 72}, 1534 (1994).
\bibitem{tilley} D.R.  Tilley and J.  Tilley, {\em Superfluidity and
Superconductivity} (Adam Hilger, Bristol 1990).  \bibitem{donnelly}
R.J.  Donnelly, {\em Quantizes Vortices in Helium II} (Cambridge
University Press, Cambridge 1991).  \bibitem{aoth} P.  Ao and D.J.
Thouless, {\em Phys.  Rev.  Lett.}  {\bf 70}, 2158 (1993).
\bibitem{thaoni} D.J.  Thouless, P.  Ao and Q.  Niu, {\em Physica}
{\bf A 200}, 42 (1993).  \bibitem{gait} F.  Gaitan, ICTP preprint
IC/94/383.  \bibitem{clem} J.R.  Clem, {\em Phys.  Rev.}  {\bf B 43},
7837 (1991).  \bibitem{calleg} A.O.  Caldeira and A.J.  Leggett, {\em
Ann.  Phys.} {\bf 149}, 374 (1983).  \bibitem{ao} Ping Ao, {\em J.
Low Temp.  Phys.}  {\bf 89}, 543 (1992).  \bibitem{niaoth} Q.  Niu,
P.  Ao and D.J.  Thouless, {\em Phys.  Rev.  Lett.}  {\bf 72}, 1706
(1994).  \bibitem{leefish} D.-H.  Lee and M.P.A.  Fisher, {\em Int.
J.  Mod.  Phys.}  {\bf B 5}, 2675 (1991).  \bibitem{minn} P.
Minnhagen, {\em Rev.  Mod.  Phys.}  {\bf 59}, 1001 (1987).
\bibitem{itzzub} C.  Itzykson and J.B.  Zuber, {\em Quantum Field
Theory} (McGraw-Hill, New York 1980).  \bibitem{guinea} F.  Guinea
and Yu.  Pogorelov, {\em Phys.  Rev.  Lett.}  {\bf 74}, 462 (1995).
\bibitem{shenoy} B.  Chattopadhyay and S.R.  Shenoy, {\em Phys.
Rev.  Lett.} {\bf 72}, 400 (1994).  \end{thebibliography}
\end{document}